   \documentclass{aa}
\usepackage{graphicx}
\usepackage{natbib}
\bibpunct{(}{)}{;}{a}{}{,}

\usepackage{txfonts}
\begin{document}
\title{Hinode/EIS observations of propagating low-frequency slow magnetoacoustic waves in fan-like coronal loops}

\author{T. J. Wang\inst{1,2}, L. Ofman\inst{1,2}, J. M. Davila\inst{2}, and J. T. Mariska\inst{3}}
     \offprints{T.J. Wang}
     \institute{Department of Physics, Catholic University of America, 620 Michigan Avenue, Washington, DC 20064, USA 
\\
     \email{wangtj@helio.gsfc.nasa.gov}
     \and
       NASA Goddard Space Flight Center, Code 671, Greenbelt, MD 20771, USA\\
     \and
       Space Science Division, Naval Research Laboratory, Washington, DC 20375, USA}

\date{Received ----; accepted ----}

\abstract
{}
{We report the first observation of multiple-periodic propagating disturbances along a fan-like coronal 
structure simultaneously detected in both intensity and Doppler shift in the Fe\,{\sc{xii}} 195 \AA\ line 
with the EUV Imaging Spectrometer (EIS) onboard $Hinode$. A new application of coronal seismology is provided 
based on this observation.}
{We analyzed the EIS sit-and-stare mode observation of oscillations using the running difference and 
wavelet techniques.}
{Two harmonics with periods of 12 and 25 min are detected. We measured the Doppler shift amplitude 
of 1$-$2 km~s$^{-1}$, the relative intensity amplitude of 3\%$-$5\% and the apparent 
propagation speed of 100$-$120 km~s$^{-1}$.}
{The amplitude relationship between intensity and Doppler shift oscillations provides convincing 
evidence that these propagating features are a manifestation
of slow magnetoacoustic waves. Detection lengths (over which the waves are visible) of the 25 min wave are 
about 70$-$90 Mm, much longer than those of the 5 min wave previously detected by TRACE. This
difference may be explained by the dependence of damping length on the wave period for thermal conduction.
Based on a linear wave theory, we derive an inclination of the magnetic field to the line-of-sight about 
59$\pm$8$^{\circ}$, a true propagation speed of 128$\pm$25 km~s$^{-1}$ and a temperature of 0.7$\pm$0.3 MK 
near the loop's footpoint from our measurements.}

\titlerunning{Propagating low-frequency slow magnetoacoustic waves in coronal loops}
\authorrunning{Wang et al.}

\maketitle

\section{Introduction}

Quasi-periodic propagating intensity disturbances in fanlike large coronal loops  
were first observed with SOHO/EIT 195 \AA\ data \citep{ber99}, and then were confirmed with TRACE 171 \AA\ 
data \citep[e.g.][]{dem00}. These disturbances with a propagation speed on the order of 150 km~s$^{-1}$ 
have been interpreted as slow magnetoacoustic waves \citep{nak00}. Detection of these waves is 
important for our understanding of the energy balance in the outer solar atmosphere 
and also highly valuable for coronal seismology \citep[e.g. reviews by][]{nak05}.

Many authors have found evidence of 3 min and 5 min oscillations propagating through 
the chromosphere and transition region into the lower corona \citep[e.g.][]{dem02,mar03, wan09}, supporting
their origin in the photospheric $p$-modes by wave leakage \citep{dep05}. In addition, there were also a 
few reports of similar propagating waves with long periods (10$-$15 min) observed in coronal loops by EIT 
\citep{ber99}, TRACE \citep{mci08}, and STEREO/EUVI \citep{mar09}, but their origin is still not clear. 
Long-period (10$-$30 min) oscillations were also observed in polar coronal holes 
\citep[e.g.][]{ban09}.

In this study, we detect two long-period (12 and 25 min) harmonics in propagating 
slow magnetoacoustic waves in fanlike coronal loops with observations from the EUV Imaging Spectrometer 
(EIS) onboard {\it Hinode}. The EIS instrument with high sensitivity and spectral resolution \citep{cul07} 
can achieve relative Doppler-shift measurements with an accuracy of less than 1 km~s$^{-1}$ 
\citep{mari08, van08, erd08}, providing us a good opportunity for detecting oscillations and waves in 
solar atmosphere. 

\begin{figure*}
\centering
\includegraphics[width=14.5cm]{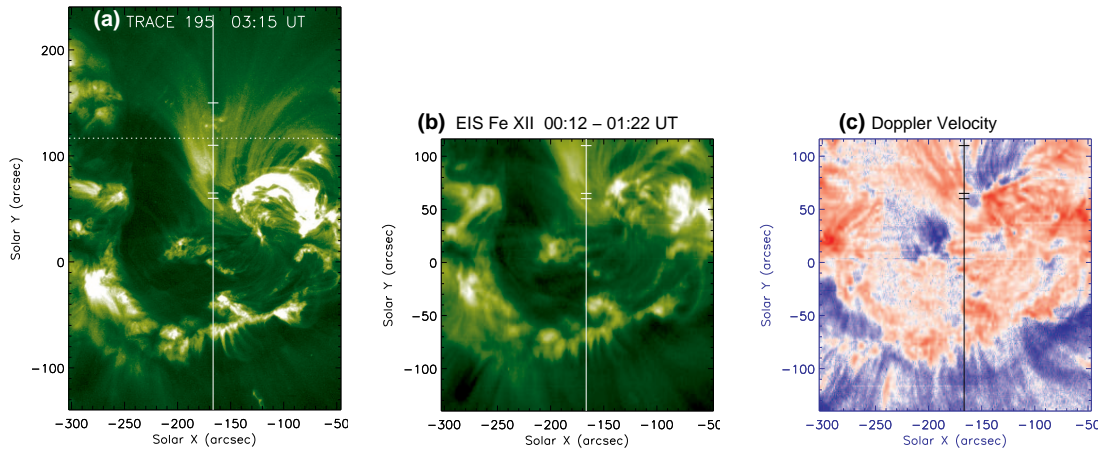}
\caption{
 (a) The TRACE $\lambda$195 bandpass image. 
(b) The intensity map in the Fe\,{\sc xii} $\lambda$\,195.12 line
from EIS. The raster observation was taken from 00:12 to  01:22 UT on 2007 February 1. (c) The Doppler 
velocity measurements. The red color represents the redshift and the blue color the blueshift with
a scale range from $-$20 km~s$^{-1}$ to $+$20 km~s$^{-1}$.  The vertical line in each plot shows
the position of the EIS 1$^{''}$ slit for the sit-and-stare observation. The short horizontal lines on the 
slit mark the positions where the oscillations are analyzed. }
\label{fgmap}
\end{figure*}

\section{Observation}
EIS has both imaging and spectroscopic 
capabilities. Its spectroscopic mode can operate in a rastering mode or a sit-and-stare mode.
In the latter case, the 1${''}$ or 2${''}$ slit is placed at a fixed location on the Sun, and repeated 
exposures are obtained while the Hinode spacecraft tracks the solar rotation.

The observations  were obtained on 2007 February 1 in AR 10940. An EIS spectroheliogram 
was taken from 00:12 to 01:22 UT with the 1${''}$ slit and covering a $256{''}\times256{''}$ region.
A sit-and-stare observation within the region was taken from 01:32 to 07:32 UT with the 1${''}\times512{''}$
slit and an exposure time of 60~s. The data cover 17 spectral windows with a cadence of about 62~s. 
This paper presents the result mainly for the Fe\,{\sc xii} $\lambda$\,195 line. 
The raw data were first processed using the standard routine {\em eis\_prep} in {\em Solar Software}
(SSW). The Fe\,{\sc xii} 195.12 \AA\ line was then fitted with a single Gaussian profile, providing 
the total intensity, Doppler shift, and line width. The effect of a weak blended line, 195.18 \AA\
\citep{you09}, is negligible since only relative Doppler shift variations are concerned in our study. 

Hinode is known to have instrumental jitter in both the $x$ and $y$ directions. 
We estimated drifts of the pointing using the SSW
routine {\it eis\_jitter} from the satellite alignment data \citep{shi07}, and find 
that the displacements of the EIS pointing are within 3${''}$ during the observation. For the 
sit-and-stare data analyzed below, the pointing drifts in the $y$-direction have been corrected.
The drifts in the $x$-direction have little effect on our result since they are slow and 
orbit-related with an amplitude less than 2${''}$ and a period of about 90 min.

\begin{figure*}
\centering
\includegraphics[width=14.5cm, height=6.3 cm]{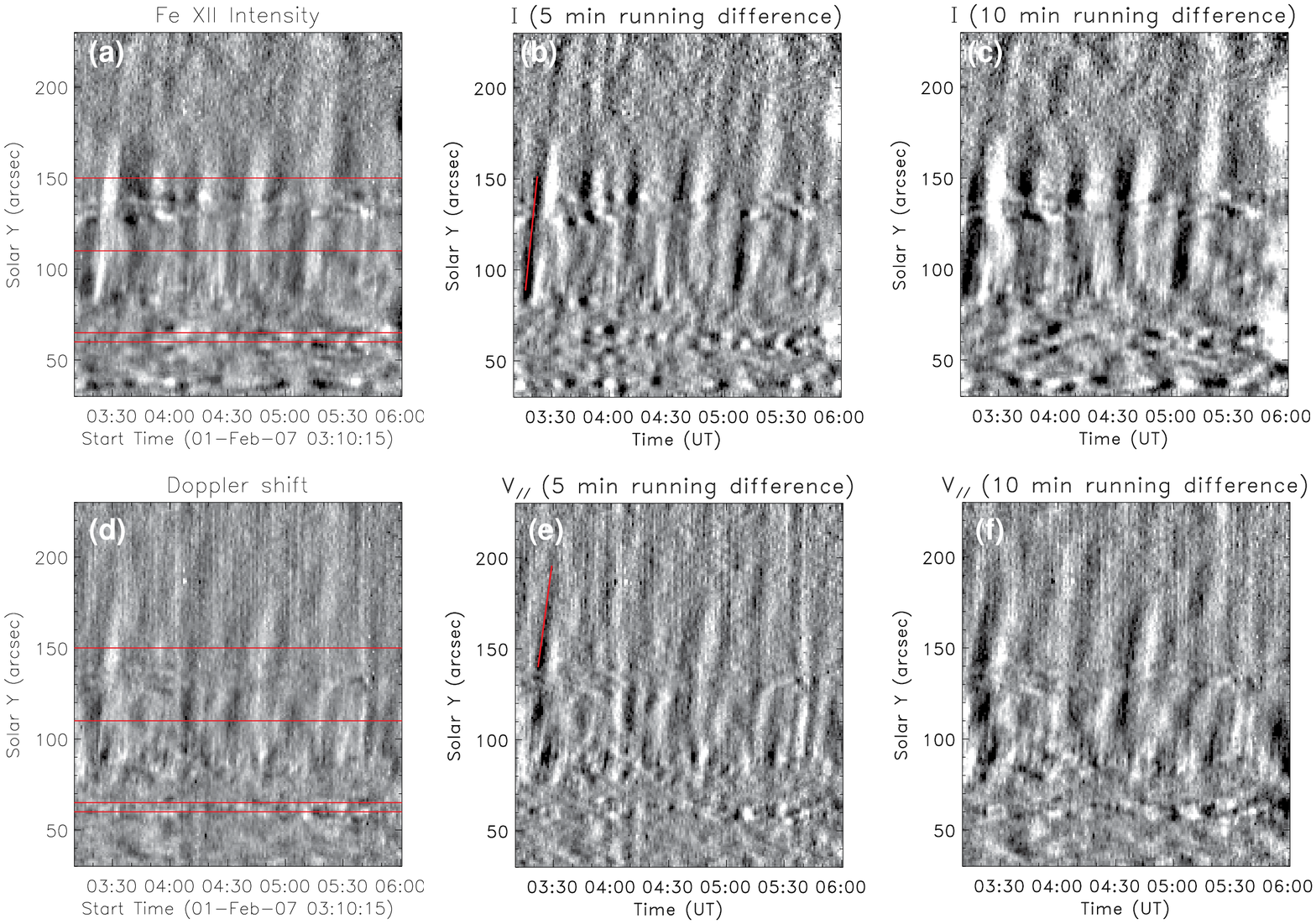}
\caption{
The upwardly-propagating waves in coronal loops observed by Hinode/EIS. (a) Time series of
relative intensity along the slit in the Fe\,{\sc xii} 
$\lambda$195.12 line. (b) and (c) The 5 min and 10 min running difference maps for intensity.  
(d) Time series of Doppler shift. Here the white color indicates the blueshift and the black color 
indicates the redshift. (e) and (f): The same as (b) and (c) but for Doppler shift. The horizontal lines  
in (a) and (d) mark three positions ($y$=150$^{''}$, 110$^{''}$, and 60$^{''}$$-$65$^{''}$), where the time
series of oscillations are analyzed (see Figs.\,\ref{fgosc1} and \ref{fgosc2}, and Fig.\,\ref{fgosc3}
(online only). The lines in (b) and (e) outline the propagation features along the slit to demonstrate 
measurements of the propagation speed.
}
\label{fgprp}
\end{figure*}

\section{Results}
A part of TRACE 195 \AA\ image recorded at 03:15 UT shows a typical fan-like set of diverging coronal loops, 
located at the northern edge of NOAA active region 10940  (Fig.\,\ref{fgmap}a). The EIS spectroheliogram 
taken at 00:12 UT only covers part of this structure (Fig.\,\ref{fgmap}b). The slit used for the sit-and-stare
observation is almost aligned along a loop. The Doppler shift map in EIS Fe\,{\sc xii} $\lambda$195 
shows that the eastern part of footpoints of the fan-like loops is dominated by redshifts, while the western 
part of footpoints is dominated by blueshifts (Fig.\,\ref{fgmap}c). The sit-and-stare mode observations 
show weak blueshifts less than 7 km~s$^{-1}$ at a region of $y$=90$^{''}$$-$$120^{''}$.  

Time-distance maps in intensity and Doppler shift for the EIS
Fe\,{\sc xii} line are shown in Fig.\,\ref{fgprp}. To enhance weak propagating intensity disturbances, 
the time series of relative intensity were normalized to the background trend at each position along the slit
by [$I(t,y_i)-I_{bg}(t,y_i)$]/$I_{bg}(t,y_i)$ (Fig.\,\ref{fgprp}a), where $I_{bg}(t,y_i)$ is the 
background trend at $y=y_i$ taken as a 20 min average smoothing of $I(t,y_i)$, while the 30 min
average smoothing gives almost no difference. The background trend was also subtracted from the time
series of Doppler shift with a similar method (Fig.\,\ref{fgprp}d). The upwardly propagating disturbances
can be seen in both intensity and Doppler shift as weak bright and dark bands with the positive gradient. 
The propagating disturbances are slightly affected by a small bright point at $y$=130$^{''}$$-$$140^{''}$. 
The footpoint of the loop (at $y$=80$^{''}$$-$$100^{''}$) appears to be characterized by the relatively 
high-frequency features, while the upper part is dominated by the relatively low-frequency features. 
In other words, the lower frequency component appears to propagate over a longer distance than the 
higher frequency one. These time-variable features can be enhanced in a running difference image, 
which was created by subtracting from each frame a frame taken some exposures later. 
The middle and right panels of Fig.\,\ref{fgprp} show the 5 min and 10 min running difference images,
which clearly reveal the presence of multiple-period modes in the propagating disturbances.
The propagation velocity transverse to the line of sight is estimated as the gradient of 
the lines outlining the evident slanting bands in the running difference images (see two examples shown 
in Figs.\,\ref{fgprp}b and \ref{fgprp}e). For nine propagation features selected for intensity and Doppler
shift separately, we measured the projected propagation speed of 118$\pm$34 km~s$^{-1}$ 
for intensity and 105$\pm$25 km~s$^{-1}$ for Doppler shift. In addition, we also examined the coordinated
TRACE data, but no corresponding propagating features are found in the constructed time-distance image 
along the EIS slit (see Fig.\,\ref{fg1trc} online). This may come from the lower sensitivity of TRACE compared to EIS.

The periods of the disturbances were investigated using a wavelet analysis method \citep{tor98}.
The Morlet wavelet was chosen for the convolution of time series.
The intensity and Doppler shift oscillations at three positions ($y$=150$^{''}$, 110$^{''}$, and 
60$^{''}$$-$$65^{''}$, denoted as oscillations 1, 2, and 3) were selected for analysis. Oscillation 1  
is located at the position of a loop with a projected distance of 
about 50 Mm away from the footpoint, near which oscillation 2 is measured. Oscillation 3 was selected for 
comparison, located at another loop intersecting the slit (Fig.\,\ref{fgmap}a). The time profiles
used for the analysis of oscillation 3 were obtained by averaging over six pixels along the slit because the
oscillation region varied slightly in the $y$-direction during the observation. In practice, we first subtracted the background trend for both intensity and Doppler shift, which was obtained by temporally smoothing 
the series. Since oscillation 1 is dominated by the lower-frequency wave component, the background is taken 
as a 20 min average smoothing. To emphasize the shorter-frequency wave component in oscillations 2 and 3, 
the background is chosen as a 10 min average smoothing.   

\begin{figure}
\centering
\includegraphics[width=8cm, height=7cm]{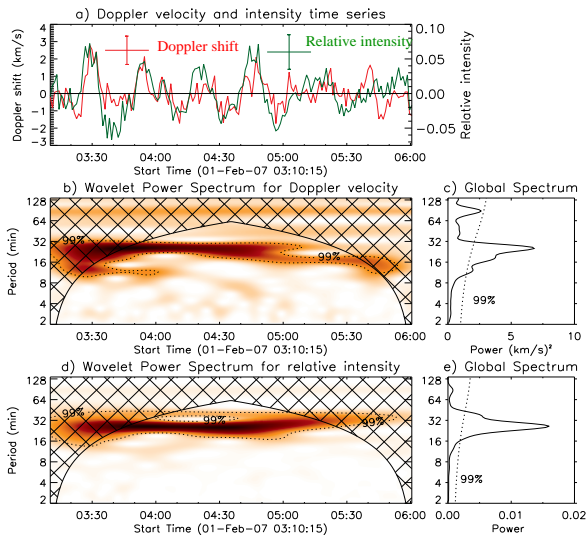}
\caption{
Wavelet analysis for oscillation 1 at $y$=150$^{''}$. (a) Time profiles of Doppler shift and relative intensity.
Here positive values for the Doppler shift represent the blueshifted emission. 
(b) The wavelet power spectrum for Doppler shift. The dark color represents high power and the dotted 
contour encloses regions of greater than 99\% confidence for a white-noise process. (c) The global wavelet spectrum 
(solid line) and its 99\% confidence level (dotted line). (d) and (e): The same as (b) and (c) but for relative
intensity.
}
\label{fgosc1}
\end{figure}

\begin{figure}
\centering
\includegraphics[width=8cm, height=6.5cm]{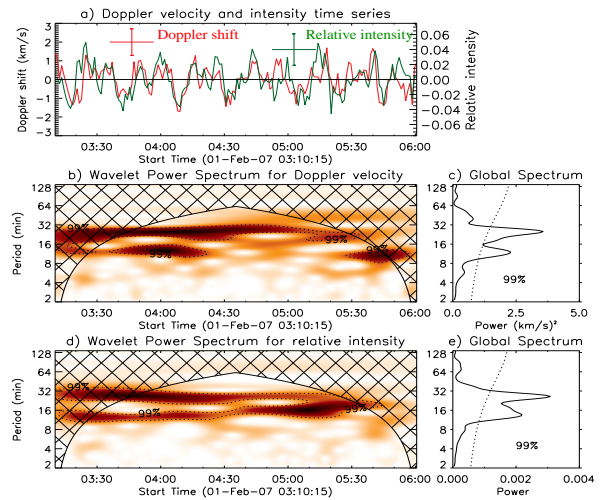}
\caption{
Wavelet analysis for oscillation 2 at $y$=110$^{''}$. The annotations are same as in Fig.\,\ref{fgosc1}. 
}
\label{fgosc2}
\end{figure}

The evolution of detrended relative intensity and Doppler shift for three oscillations is shown in panels 
(a) of Figs.\,\ref{fgosc1} and \ref{fgosc2}, and of Fig.\,\ref{fgosc3} (online only). 
There is a good in-phase relationship between the intensity and
Doppler velocity, which is consistent with the signature as predicted by linear MHD wave theory for 
an upwardly propagating slow magnetoacoustic wave in coronal loops. The wavelet spectrum and the global
wavelet spectrum are shown in panels (b)-(e) of Figs.\,\ref{fgosc1} and \ref{fgosc2}, and of Fig.\,\ref{fgosc3}
(online only). Cross-hatched regions
in the wavelet spectrum indicate the ``cone of influence", where edge effects become important due to
the finite length of time series. The global wavelet spectrum is the average of the wavelet power over time
at each oscillation period. 

For oscillation 1 the wavelet spectra for both intensity and Doppler shift 
shows strong power at the period in a range of 20$-$30 min over a duration of 6 cycles. From the global 
wavelet spectra, the oscillation period for Doppler shift is measured to be 25.6$\pm$6.3 min and the
period for intensity is 26.2$\pm$4.9 min. The uncertainties are taken as the half FWHM of the peak.  
The average maximum amplitude for Doppler shift is 1.6$\pm$0.5 km~s$^{-1}$
and for relative intensity is (4.9$\pm$1.5)\%, where the errors are the standard deviation of 
absolute peak values. As a comparison, we also examined the case with a subtraction of 
the 30 min smoothing background (see Fig.\,\ref{fg1sm30} online), and find that the power 
of the 26 min band shows almost no change, while the power of the 88 min band for Doppler shift rises
above the 99\% confidence level, which is caused by the orbital effect.

Oscillation 2 clearly shows that most of the wavelet power is concentrated within two period bands ranging 
from 8$-$16 min and from 20$-$30 min, indicating that the waves near the footpoint consist of multiple harmonic modes. 
The two periods are measured to be 11.9$\pm$2.8 min and 25.1$\pm$7.0 min for Doppler shift, and 13.5$\pm$4.7 min 
and 26.2$\pm$ 5.0 min for intensity. The average maximum amplitude for Doppler shift is 1.3$\pm$0.3 km~s$^{-1}$
and for intensity is (3.8$\pm$0.6)\%. Oscillation 3 shows most of the wavelet powers for Doppler 
shift and intensity concentrated in a range of 8$-$14 min. The measurements of periods and amplitudes for
oscillations 1$-$3 are summarized in Table\,\ref{tabpar} (online only).

We show a new application of coronal seismology based on interpreting our observation in terms of
slow magnetoacoustic waves. Assuming that the coronal field, along which the waves propagate, has an 
inclination, $\phi$, to the line of sight, we have
\begin{equation}
{\rm sin}{\phi}=\frac{V_p}{c_s}, \label{eqsn}
\end{equation} 
where $V_p$ is the projected propagation speed of the waves and $c_s$ the sound speed, which is a good
approximation to the phase speed of slow magnetoacoustic waves under a typical coronal condition of 
low-$\beta$ plasma.  The linearized 
continuity equation gives $\rho^{\prime}/\rho_0=V^{\prime}/c_s$, where $\rho_0$ is the undisturbed loop
density, and $\rho^{\prime}$ and $V^{\prime}$ the density and velocity perturbations, respectively. 
Considering $I^{\prime}/I_0\sim2\rho^{\prime}/\rho_0$ and ${\rm cos}{\phi}=V_{\parallel}/V^{\prime}$, 
where $I^{\prime}/I_0$ is the relative amplitude for intensity disturbances and $V_{\parallel}$ 
the Doppler shift amplitude, we obtain
\begin{equation} 
 {\rm cos}{\phi}=\frac{2V_{\parallel}}{(I^{\prime}/I_0)c_s}. \label{eqcs}
\end{equation}
From Eqs.\,(\ref{eqsn}) and (\ref{eqcs}) we derive the inclination angle,
\begin{equation}
\phi={\rm tan}^{-1}\left[\frac{1}{2}V_p\left(\frac{I^{\prime}/I_0}{V_{\parallel}}\right)\right].  \label{eqph}
\end{equation}
Note that Eq.\,(\ref{eqph}) also holds in the condition with a steady flow in the loop.
We take $V_p$=110$\pm$30 km~s$^{-1}$, the mean of the projected propagation speeds measured for intensity and 
Doppler shift. By assuming that the inclination of the coronal field is constant over the detection length
and taking the quantity of $(I^{\prime}/I_0)/V_{\parallel}$ to be the mean value of the
measurements for oscillations 1 and 2, we obtain $\phi=59^{\circ}\pm8^{\circ}$ from Eq.\,(\ref{eqph}).
Then with Eq.\,(\ref{eqsn}) the sound speed is estimated to be 128$\pm$25 km~s$^{-1}$. From $c_s=0.152T^{1/2}$ 
km~s$^{-1}$ we estimate the average temperature of the loop near the footpoint about 0.7$\pm$0.3 MK. This result
is in good agreement with what is obtained from observations of 3D propagation of similar waves with STEREO/EUVI
\citep{mar09}.

\section{Discussion}
We have reported on the first observation of the multiple-period (12 and 25 min) propagating disturbances 
along a fan-like coronal structure in both intensity and Doppler shift in the Fe\,{\sc xii} 195 \AA\ line with Hinode/EIS. 
The phase and amplitude relationships between the intensity and Doppler shift oscillations provide 
convincing evidence that they are a propagating slow magnetoacoustic wave.

With high cadence (15 s) EIT 195 \AA\ observations, \citet{ber99} first detected the propagating 
intensity disturbances in fanlike coronal loops. They noticed that these disturbances are associated 
with recurrent footpoint brightenings, appearing as transient flows ejected periodically. In our case 
the propagating features are detected in both intensity and Doppler shift. The small Doppler
shift amplitude of several km~s$^{-1}$ is consistent with the intensity amplitude as expected for 
a slow magnetoacoustic wave, thus, interpreting these oscillations in terms of propagating 
slow magnetosonic waves is more appropriate. In addition, the apparent propagation
speed of the oscillations $>$100 km~s$^{-1}$ is an order of magnitude higher than any flow speeds detected 
in this region. 

In our case the dominant oscillations have periods of about 25 min, much longer than those observed
in the TRACE loops of around 5 min \citep{dem02}. It is not clear why the longer periods reported here 
were not observed in TRACE loops. One reason for \citet{dem02} 
detecting only the 5 min oscillations could be attributed to the selection criterion they used, 
where a running difference technique with time lag of 90 s would eliminate the long-period components 
that may exist in some cases. \citet{dep05} suggest that 5 min waves in coronal loops come 
from the leakage of $p$-modes through the chromosphere tunneled by highly inclined magnetic fields. 
However, the origin of 20$-$30 min waves is difficult to explain by such a mechanism because their
period is much longer than the cutoff period of the slow magnetoacoustic waves in the chromosphere 
\citep{bel77}. Therefore, another explanation is required, e.g., possibly the coupling with the photospheric
$f$-mode caused by local convective motions. 

The detection length is defined as the distance along the loop over which the intensity disturbances 
can be observed. TRACE observations show that the waves are damped quickly with a detection length 
typically in the range of 3$-$23 Mm \citep{dem02}. Theoretical and numerical studies show that 
thermal conduction can play an important role in the quick damping of slow magnetosonic waves 
in coronal loops \citep[e.g.][]{ofm02, dem04, kli04}. In our case the detection length is measured 
to be about 69 Mm for intensity and about 87 Mm for Doppler velocity, much longer than 
in the TRACE case. This difference may be explained by the dependence of the damping length on the
wavelength and the temperature of the ambient plasma \citep{kli04}. Thus, the longer period waves 
reported here have longer wavelength than in previous observations, resulting in weaker damping by 
thermal conduction. In addition, the present loop appears cooler ($\sim$ 0.7 MK as estimated in the
previous section) than previously observed loops (1.0$-$1.5 MK for TRACE loops).

\acknowledgements
 {\it Hinode} is a Japanese mission developed and launched by ISAS/JAXA in partnership with
NAOJ, NASA, and STFC (UK). Additional operation support is provided by ESA and NSC (Norway).
The authors are grateful to Dr. Harry Warren for his planning of EIS observations. The work of LO and 
TJW was supported by NRL grant N00173-06-1-G033. LO was also supported by NASA grant NNG06GI55G.
The authors also thank the referee, Dr. Dipankar Banerjee, for his constructive comments and suggestions.

\begin{newpage}
\appendix 
\section{Online Materials}

\begin{figure}
\centering
\includegraphics[width=8cm]{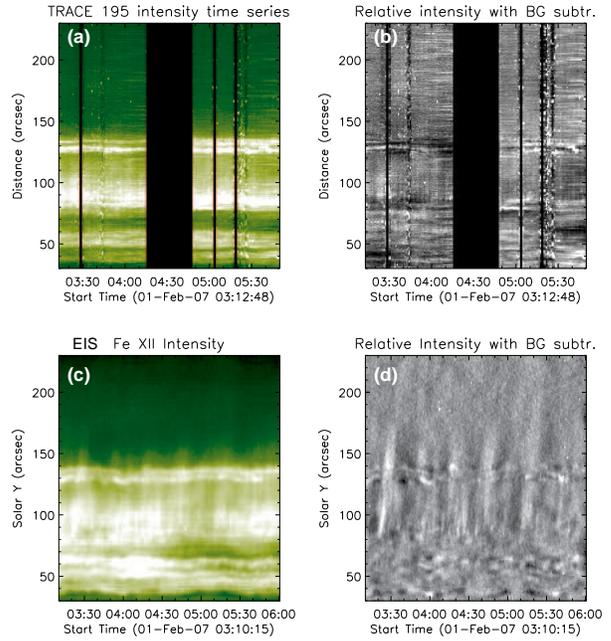}
\caption{ 
Comparisons between EIS and TRACE observations. (a) Constructed TRACE 195 \AA\ time-distance image along the EIS slit. 
(b) Same as (a) but with the background subtraction, where the background is taken as the average over time at each
position along the cut. (c) The EIS Fe\,{\sc xii} intensity time series. (d) Same as (c) but
with the background subtraction.
}
\label{fg1trc}
\end{figure}

\begin{figure}
\centering
\includegraphics[width=8cm]{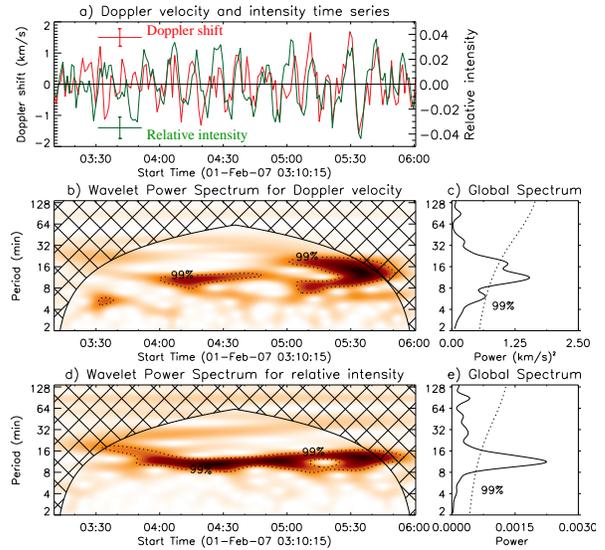}
\caption{ 
Wavelet analysis for oscillation 3 averaged over $y$=60$^{''}$$-$65$^{''}$. The annotations are same as 
in Fig.\,\ref{fgosc1}.
}
\label{fgosc3}
\end{figure}

\begin{figure}
\centering
\includegraphics[width=8cm]{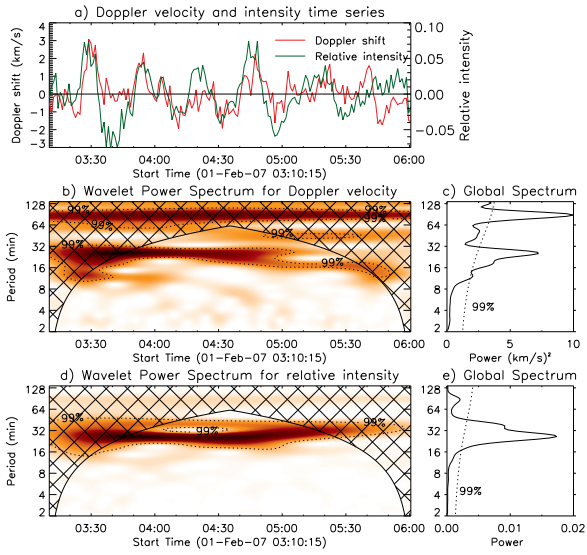}
\caption{ 
Wavelet analysis for oscillation 1 at y=150${''}$, which is the same as Fig.\,\ref{fgosc1} but with a subtraction of 
the background taken as a 30 min average smoothing. 
}
\label{fg1sm30}
\end{figure}

\begin{table}
 \begin{center}
\caption{Measurements of the oscillations in fanlike loops found in the EIS sit-and-stare data in 
the Fe\,{\sc xii} (195\AA) line$^{\mathrm{a}}$.}
\label{tabpar}
\begin{tabular}{lcccc}
\hline\hline
 Oscillation & $P_V$ & $P_I$ & $A_V$ & $A_I$ \\
  No.        & (min) & (min) & (km~s$^{-1}$) & (\%) \\
\hline
1 & 25.6$\pm$6.3 & 26.2$\pm$4.9 & 1.6$\pm$0.5 & 4.9$\pm$1.5\\
\hline
2$^{\mathrm{b}}$ & 25.1$\pm$7.0 & 26.2$\pm$5.0 & 1.3$\pm$0.3 & 3.8$\pm$0.6\\
  & 11.9$\pm$2.8 & 13.5$\pm$4.7 & ---- &  ---- \\
\hline
3 & 11.2$\pm$3.8 & 11.2$\pm$2.9 & 1.0$\pm$0.3 & 2.8$\pm$0.7\\
\hline
\end{tabular}
\end{center}
\begin{list}{}{}
\item[$^{\mathrm{a}}$] $P_V$ and $P_I$ are periods and $A_V$ and $A_I$ are amplitudes for the Doppler shift 
and relative intensity oscillations, respectively.
\item[$^{\mathrm{b}}$] Two harmonic components are found for oscillation 2.
\end{list}
\end{table}

\end{newpage}

\end{document}